\documentclass[a4paper,11pt]{article}

\usepackage{pos}
\usepackage{color}
\usepackage{epsfig,palatino}
\usepackage{mathrsfs}
\usepackage[T1]{fontenc}
\usepackage{graphicx}
\usepackage{ulem}
\usepackage{xcolor}
\usepackage{wasysym}
\usepackage{varioref}
\usepackage[english]{babel}
\usepackage{pgf,tikz}
\usetikzlibrary{
  automata,
  calc,
  decorations.markings,
  shapes,
  petri,
  positioning
}
\tikzset{%
  transition/.style={rectangle,minimum size=6mm,draw},
  place/.style={circle,minimum size=6mm,draw},
  database/.style={
    minimum width=2cm,minimum height=1cm,cylinder,
    shape border rotate=90,aspect=0.2,draw
  },
  ->-/.style={
    decoration={
      markings,
      mark=at position #1 with {\arrow{>}}
    },
    postaction={decorate}
  },
  ->-/.default=0.5
}

\makeindex

\newcommand{\setQ}[0]{\mathbb{Q}}

\def\d{{\rm d}}

\usepackage{subcaption}
\usepackage{xspace}

\newcommand{\GPISpace}{\textsc{GPI-Space}\xspace}

\usepackage[autostyle=true]{csquotes}

\theoremstyle{plain}

\title{Module Intersection for the Integration-by-Parts Reduction of
  Multi-Loop Feynman Integrals }
\ShortTitle{Module Intersection}
\author[a,b]{Dominik Bendle}
\author*[a]{Janko B\"ohm}
\author[a]{Wolfram Decker}
\author[c,d]{Alessandro Georgoudis}
\author[b]{Franz-Josef~Pfreundt}
\author[b]{Mirko~Rahn}
\author*[e,f]{Yang Zhang}
\affiliation[a]{ Department of Mathematics, Technische Universit\"at
        Kaiserslautern,\\ 67663 Kaiserslautern, Germany}
\affiliation[b]{Fraunhofer Institute for Industrial Mathematics ITWM,\\ Fraunhofer-Platz 1, 67663 Kaiserslautern, Germany}
\affiliation[c]{Laboratoire de physique de l'Ecole normale superieure, ENS, Universit\'e PSL, CNRS, Sorbonne Universit\'e, Universit\'e Paris-Diderot, Sorbonne Paris Cit\'e,\\
24 rue Lhomond, 75005 Paris, France}
\affiliation[d]{Institut de Physique Th\'eorique, CEA, CNRS, Universit\'e
Paris-Saclay,\\ F-91191 Gif-sur-Yvette cedex, France}
\affiliation[e]{Peng Huanwu Center for Fundamental Theory,\\ Hefei, Anhui 230026, China}
\affiliation[f]{Interdisciplinary Center for Theoretical Study, University of Science
and Technology of China,\\ Hefei, Anhui 230026, China}
\emailAdd{bendle@rhrk.uni-kl.de}
\emailAdd{boehm@mathematik.uni-kl.de}
\emailAdd{decker@mathematik.uni-kl.de}
\emailAdd{alessandro.georgoudis@phys.ens.fr}
\emailAdd{franz-josef.pfreundt@itwm.fraunhofer.de}
\emailAdd{mirko.rahn@itwm.fraunhofer.de}
\emailAdd{yzhphy@ustc.edu.cn}

\abstract{
In this manuscript, which is to appear in the proceedings of
the conference ``MathemAmplitude 2019'' in Padova, Italy, we provide
an overview of the module intersection method for the the integration-by-parts (IBP) reduction of
  multi-loop Feynman integrals. The module intersection method, based on
  computational algebraic geometry, is a highly efficient way of
  getting IBP relations without double propagator or with a bound
  on the highest propagator degree. In this manner, trimmed IBP
  systems which are much shorter than the traditional ones can be
  obtained. We apply the modern, Petri net  based, workflow management system
  \textsc{GPI-Space} in combination with the computer algebra system \textsc{Singular} to solve the trimmed IBP system via interpolation
  and efficient parallelization. We show, in particular,  how to use
  the new plugin feature of \textsc{GPI-Space} to manage a global
  state of the computation and to efficiently handle mutable data. Moreover, a
  Mathematica interface to generate IBPs with restricted propagator
  degree, which is based on module intersection, is
  presented in this review.
}
\FullConference{MathemAmplitudes 2019: Intersection Theory \& Feynman Integrals\\18-20 December 2019\\	
Padova, Italy}
  
\begin{document}

\maketitle

\section{Introduction}
In the past years, several new developments have been achieved for multi-loop scattering amplitudes computation. For example, particular interest has been put into tackling the $2\rightarrow 3$ scattering processes at the next-to-next-to-leading order \cite{Badger:2013gxa,Gehrmann:2015bfy,Badger:2017jhb,Abreu:2017hqn,Abreu:2018aqd,Abreu:2018jgq,Boels:2018nrr,Gehrmann:2018yef,
Badger:2018enw,Abreu:2018zmy,Chicherin:2018yne,Chicherin:2019xeg,Abreu:2019rpt,Abreu:2019odu,
Hartanto:2019uvl}. These theoretical inputs are needed for analyzing data coming from the LHC collider, especially for precise background processes.

In obtaining these results, the study of Feynman integrals and their properties is still a fundamental tool to achieve progress. In this context, generating integration-by-parts (IBP) identities \cite{CHETYRKIN1981159}  is an important step for computing amplitudes and Feynman integrals themselves, using the method of differential equations  \cite{Kotikov:1990kg,Kotikov:1991pm,Bern:1993kr,Remiddi:1997ny,Gehrmann:1999as,Henn:2013pwa,%
Papadopoulos:2014lla,Lee:2014ioa,Ablinger:2015tua,Papadopoulos:2015jft,Liu:2017jxz}. 
As such computations are the current bottleneck for most of the needed processes, there have been different approaches that try to bypass \cite{Dixon:2011pw,Dixon:2013eka,Dixon:2014iba,Caron-Huot:2016owq,Dixon:2015iva,Dixon:2016nkn,Chicherin:2017dob,Caron-Huot:2019vjl} or to simplify this step by utilizing numerical methods \cite{Ita:2015tya,Abreu:2017xsl,Badger:2018enw, Abreu:2018zmy, Abreu:2019rpt, Badger:2019djh}. (See also ref.~\cite{Mastrolia:2018uzb,Frellesvig:2019uqt,Frellesvig:2019kgj,Frellesvig:2020qot,Liu:2017jxz,Liu:2018dmc,Guan:2019bcx,Zhang:2018mlo,Wang:2019mnn} for integral reduction not directly applying IBPs.)

The development of new IBP algorithms, based mostly on ideas and tools coming from algebraic geometry, is the main topic of this review.
IBP relations are derived from the integration of a total derivative. By generating enough of these identities it is possible to express each integral appearing in the amplitude as a linear combination of a finite basis of master integrals (MI) using Gaussian elimination and the Laporta algorithm \cite{Laporta:2001dd}. There exist several available software packages, both private and public, to obtain such relations \cite{Smirnov:2008iw,Smirnov:2014hma, Smirnov:2019qkx,
  Maierhoefer:2017hyi, Maierhofer:2018gpa, vonManteuffel:2012np,Klappert:2020nbg}. The method presented here, relies on a different approach, where firstly the a priory knowledge of  basis of MI is needed. To obtain such a basis, different methods can be used \cite{Lee:2013hzt,Georgoudis:2016wff}. Recently it has been discovered how a careful choice of the basis can greatly simplify the reduction process \cite{Bendle:2019csk,Usovitsch:2020jrk,Smirnov:2020quc,Boehm:2020ijp} by generating simpler IBP coefficients. A second step is to construct, using the Baikov representation and the Laplace expansion \cite{Boehm:2017wjc}, a canonical set of IBPs. It is then possible to further restrict the number of these identities by requiring that no integral with double propagators appears \cite{Gluza:2010ws,Schabinger:2011dz,Larsen:2015ped}. This is achieved through a module intersection computation \cite{Boehm:2018fpv}. 
An extra step is then to construct a spanning set of cuts \cite{Larsen:2015ped}, further reducing the problem as it is possible to divide the integrals contributing to the IBP identities into different families and treat them separately\footnote{It is also possible to singularly nullify the integrals appearing in the reduction process \cite{Chawdhry:2018awn}.}. 
Finally, besides reducing the number of identities generated, it is important to have an efficient Gaussian elimination to solve the associated linear system. The most efficient method for performing this task is to utilize rational reconstruction and interpolation \cite{vonManteuffel:2014ixa,Peraro:2016wsq,Klappert:2020nbg, Peraro:2019svx}, where by sampling different numerical points one can obtain the analytical form of the IBP coefficients. In our implementation of this idea, we rely on the \textsc{Singular}-\textsc{GPI-Space} framework \cite{BDFPRR} for massively parallel computation in computer algebra, which combines the computer algebra system \textsc{Singular} \cite{singular} with the workflow management system \textsc{GPI-Space} \cite{GPI}, and thus allows us to massively parallelize the reduction and interpolation part of the algorithm. This framework relies on Petri nets to coordinate the computation in \textsc{GPI-Space}, while \textsc{Singular} is used as the computational back end. The Petri net formalism used in \textsc{GPI-Space} is an extension of the basic idea, which allows tokens in the net to be complex data structures. Modifying our original implementation in the \textsc{Singular}-\textsc{GPI-Space} framework, we illustrate the use of the new plugin feature of \textsc{GPI-Space} to extend the Petri net formalism to manage a global state of the computation and integrate mutable data into the formalism.
The IBP reduction method described here has been successfully used to obtain reduction for the non-planar 5-point topologies appearing at 2-loop order \cite{Boehm:2018fpv,Bendle:2019csk}. 

In this review paper, we also present a Mathematica interface for generating IBPs without double propagators (or IBPs with restricted propagator degree), via our module intersection method, see
\begin{center}
\url{https://bitbucket.org/yzhphy/module-intersection/src}
\end{center}

This review is organized as follows: In Section \ref{sec modint}, we present the intersection algorithm used to generate the IBP system. In Section \ref{sec gpi}, we describe our Gaussian reduction approach, introducing the \textsc{GPI-Space} system and the Petri net formalism to formulate our workflow. We then present some examples and finish with some outlook on possible future directions.

\section{Module Intersection}\label{sec modint}
In this section, we explain the module intersection method of generating a trimmed IBP system.

We consider the Baikov representation of Feynman integrals.

\begin{align}
I[\alpha_1,\ldots,\alpha_m]  \hspace{0.5mm}&=\hspace{0.5mm}  C_E^L \hspace{0.7mm} U^\frac{E-D+1}{2} \hspace{-1mm}
\int_\Omega \d z_1 \cdots \d z_m P(z)^\frac{D-L-E-1}{2} \frac{1}{z_1^{\alpha_1} \cdots z_m^{\alpha_m}},
\label{eq:Baikov_representation}
\end{align}
where $P(z)$ is the Baikov polynomial
\begin{equation}
  P=\det G\left(
    \begin{array}{cccccc}
      l_1,& \ldots &l_L, &p_1, & \ldots & p_E \\
l_1,& \ldots &l_L, &p_1, & \ldots & p_E
    \end{array}
\right) \,.
\end{equation}
Here $U$ and $C_E^L$ are the Gram determinant and constant factor, which are irrelevant for the IBP reduction. The polynomial $P$ vanishes on the boundary of integration $\partial \Omega$.

In the Baikov representation, an IBP relation reads
\begin{align}
  0&=\int \d z_1 \cdots \d z_m \sum_{i=1}^m\frac{\partial}{\partial
     z_i} \bigg(a_i(z) P^\frac{D-L-E-1}{2} \frac{1}{z_1^{\alpha_1} \cdots
     z_m^{\alpha_m}}  \bigg),
\label{IBP_Baikov}
\end{align}
where the $a_i(z)$ are polynomials in the Baikov variables $z=(z_1,\ldots,z_m)$. Expanding the total derivative above, we get an IBP relation of Feynman integrals. Note that the resulting relation may not match our demands: the derivative of $P$ provides dimensionally shifted integrals, and for the traditional IBPs there is no control of the propagator power increase.  However, it is easy to meet these demands by adding constraints on the $a_i(z)$. Let $\mathcal S$ be a subset of $\{1,\ldots,m\}$, for which the propagator indices are supposed to be constrained: 
\begin{gather}
  \bigg(\sum_{i=1}^m a_i(z) \frac{\partial P}{\partial z_i} \bigg)+ b(z)
  P=0\,,
\label{module1}
\\
  a_i(z) = b_i(z) z_i \,,\quad i\in \mathcal S,
\label{module2}
\end{gather}
where $b(z)$ and the $b_i(z)$ are all polynomials in the Baikov variables. The second equation is to make sure that the resulting IBP relation does not increase the propagator power $\alpha_i$ for $i\in \mathcal S$. These are the so-called syzygy equations. 

The first syzygy equation \eqref{module1} can be easily solved by the Laplacian expansion of a symmetric matrix \cite{Ita:2015tya}, or the canonical IBPs converted to Baikov form  \cite{Ita:2015tya}. The solution set is a polynomial module called $M_1$ whose generators are at most linear in the Baikov variables. This step takes almost no computing time.

The module solving the syzygy equations \eqref{module2} clearly has the trivial generating set $a_i=z_i$, $b_i=1$, $i\in \mathcal S$. We call this solution set $M_2$. So without much computing efforts,  we get $M_1$ and $M_2$, the solution sets for the syzygy equations \eqref{module1} and \eqref{module2}, individually. 

 The main computational goal is then to get,
\begin{equation}
  M_1 \cap M_2\,.
\label{intersection}
\end{equation}
This module intersection is obtained via a module Gr\"obner
basis in a position-over-term ordering \cite{Boehm:2018fpv}.

Although the generators of $M_1$ and $M_2$ contain at most linear polynomials in the Baikov variables, the intersection computation needs the following technique to finish efficiently:
\begin{itemize}
\item {\it Localization}. Let the kinematic parameters (Mandelstam and mass parameters) be $(c_1,
\ldots ,c_t)$. Instead of the obvious computation in the ring $ R =\mathbb{Q}(c_1,
\ldots ,c_t)[ z_1,\ldots, z_m]$, we do the computation in the ring $ R'=\mathbb{Q}[ z_1, \ldots , z_m, c_1, \ldots c_t ,]$ with the block ordering
\begin{equation}
z_1, \ldots , z_m \succ  c_1, \ldots c_t .
\end{equation}
This amounts to treat parameters in the same way as the Baikov variables. This technique is essential for the multivariate computations to finish. After the Gr\"obner basis computation in $R'$, we map the result back to $R$ and remove redundant generators.
\item The use of a {\it degree bound} for the intersection computation. Usually, we do not need the full generating set of the intersection $M_1\cap M_2$. So heuristically, we can apply a degree bound on the intersection computation to reduce the computation time. In practice, this is done via the ``degBound'' option in {\sc Singular}. A posteriori, we verify that the degree bound was chosen large enough.
\end{itemize}

To this proceedings article, we attach a \textsc{Mathematica} interface to {\sc Singular} for the module intersection computation and for generating constricted IBP relations. A pure {\sc Singular} version of this will be available in the future as a {\sc Singular} library.

\section{Efficient Gaussian Elimination with GPI-Space}\label{sec gpi}
In this section we discuss our parallel implementations in the workflow management system \GPISpace \cite{GPI} in combination with the computer algebra system \textsc{Singular} \cite{singular} using the framework developed in \cite{BDFPRR}. In the computation of Feynman integrals, \GPISpace is used to parallelize the Gaussian reduction of the linear system which is derived from the IBP identities. This is achieved by splitting the large reduction job into numerous small ones by passing to \enquote{semi-numeric} computations, which are then recombined using interpolation.

In the following, we will give a short overview of \GPISpace and Petri nets, which are used to formulate parallel workflows. We will then describe the Petri nets used in the computation of Feynman integrals. For more details, also refer to  \cite{bendle:master}.

\subsection{\fontfamily{cmr}\GPISpace and Petri Nets}

\GPISpace is a workflow management system developed by the Fraunhofer Institute for Industrial Mathematics (ITWM) and is comprised of three components:
\begin{itemize}
  \item The \textit{distributed run-time system} (DRTS) initializes and manages workers in accordance with the available computing resources, and allocates computing jobs as they become available.
  \item The \textit{workflow engine} (WE) tracks the state of the workflow. Jobs available for execution are identified and, together with their input data, sent to the DRTS for scheduling.
  \item A \textit{virtual memory layer} allows communication and data sharing between different machines managed by \GPISpace.
\end{itemize}

\textit{Petri} nets serve as the top level description of workflows for the \GPISpace engine. They can be defined as directed bipartite paths where the nodes are divided into \textit{places} and \textit{transitions}, which model the data that is passed between procedures, and data processing algorithms, respectively. Depending on the orientation of the directed vertex between places and transitions, places may be \textit{input} or \textit{output places} of a given transition. Places may hold \textit{tokens}, which represent the data, and can be consumed or produced by transitions. The association of tokens to places at a given time is called a \textit{marking} of the Petri net.

\begin{figure}[htbp]
  \centering
  \begin{subfigure}{.35\textwidth}
    \centering
    \begin{tikzpicture}[node distance=1.4cm]
      \node [place]      (in)                {};
      \node [transition] (f)   [right of=in] {$t^2$} edge [pre] (in);
      \node [place]      (out) [right of=f]  {} edge [pre] (f);
    \end{tikzpicture}
    \caption{Data parallelism}
    \label{fig:petri:datapar}
  \end{subfigure}
  \begin{subfigure}{.64\textwidth}
    \centering
    \begin{tikzpicture}[node distance=1.4cm]
      \node [place]      (in) {};
      \node [transition] (s)  [right of=in] {$s$} edge [pre] (in);

      \node [place]      (su) [above right of=s] {}    edge [pre] (s);
      \node [transition] (f)  [right of=su]      {$f$} edge [pre] (su);
      \node [place]      (l)  [right of=f]       {}    edge [pre] (f);

      \node [place]      (sd) [below right of=s] {}    edge [pre] (s);
      \node [transition] (g)  [right of=sd]      {$g$} edge [pre] (sd);
      \node [place]      (r)  [right of=g]       {}    edge [pre] (g);

      \node [transition] (j)  [below right of=l] {}    edge [pre] (l) edge [pre] (r);
      \node [place]      (o)  [right of=j]       {}    edge [pre] (j);
    \end{tikzpicture}
    \caption{Task parallelism}
    \label{fig:petri:taskpar}
  \end{subfigure}
  \caption{Minimal Petri nets illustrating parallelisms
  }
  \label{fig:petriEx}
\end{figure}
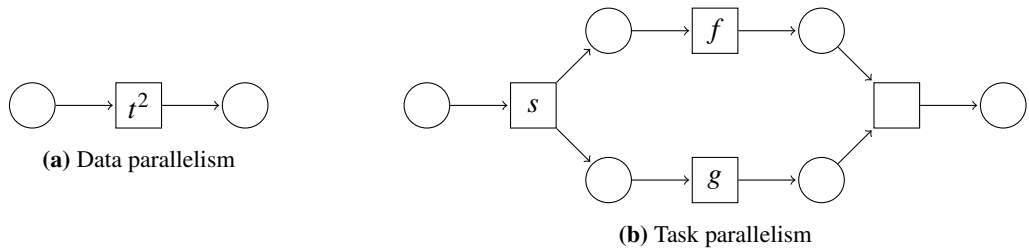

If all input places of a transition hold at least one token, the transition is called \textit{enabled} and may \textit{fire}, that is, one token is consumed from each input place, and each output place receives a new token. If we have a marking $M$ of a Petri net and obtain a new marking $M'$ by firing the transition $t$, we write $\smash{M \overset t\longrightarrow M'}$. If $t$ can fire multiple times, producing a sequence of markings
\[
  M \overset t\longrightarrow M^{(1)} \overset t\longrightarrow \cdots \overset t\longrightarrow M^{(n)}
\]
\begin{figure}[tbp]
  \centering
  \begin{tikzpicture}[node distance=1.4cm]
    \begin{scope}
      \node [place,tokens=2] (in)                {};
      \node [place,tokens=3] (in2) [above of=in] {};
      \node [transition]     (f)   [right of=in,yshift=0.7cm] {$t$}
        edge [pre] (in)
        edge [pre] (in2);
      \node [place]          (out) [right of=f]  {} edge [pre] (f);
    \end{scope}

    \begin{scope}[xshift=7cm]
      \node [place]          (in')                {};
      \node [place,tokens=1] (in2') [above of=in'] {};
      \node [transition]     (f')   [right of=in',yshift=0.7cm] {$t$}
        edge [pre] (in')
        edge [pre] (in2');
      \node [place,tokens=2] (out') [right of=f']  {} edge [pre] (f');
    \end{scope}

    \draw [->,thick] ([xshift=5mm]out -| out) -- ([xshift=-5mm]f' -| in')
      node[above=1mm,midway,text centered] {$t^2$};
  \end{tikzpicture}
  \caption{Transition firing until it is disabled. 
  }
  \label{fig:firing}
\end{figure}
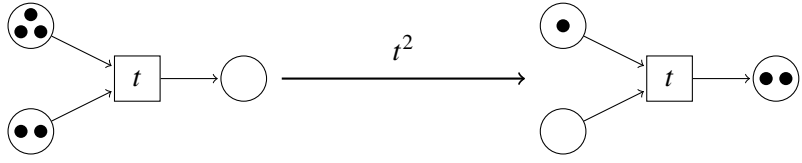
by a sequential execution of the firings,  the transition $t$, in fact, can fire in parallel for each tuple of available input tokens.
This form of parallelisim offered by Petri nets is referred to by the notation $\smash{M \overset{t^n}\longrightarrow M^{(n)}}$ and is called \textit{data parallelism}. Examples of data parallelism are shown in Figures~\ref{fig:firing} and \ref{fig:petri:datapar}. In figures, we depict transitions as boxes and places as circles. The flow relations between places and transitions is given by arrows. Of course, this parallelism extends to multiple transitions as well. If several transitions are enabled, they can fire at the same time. This is called \textit{task parallelism} and is illustrated in Figure~\ref{fig:petri:taskpar}.

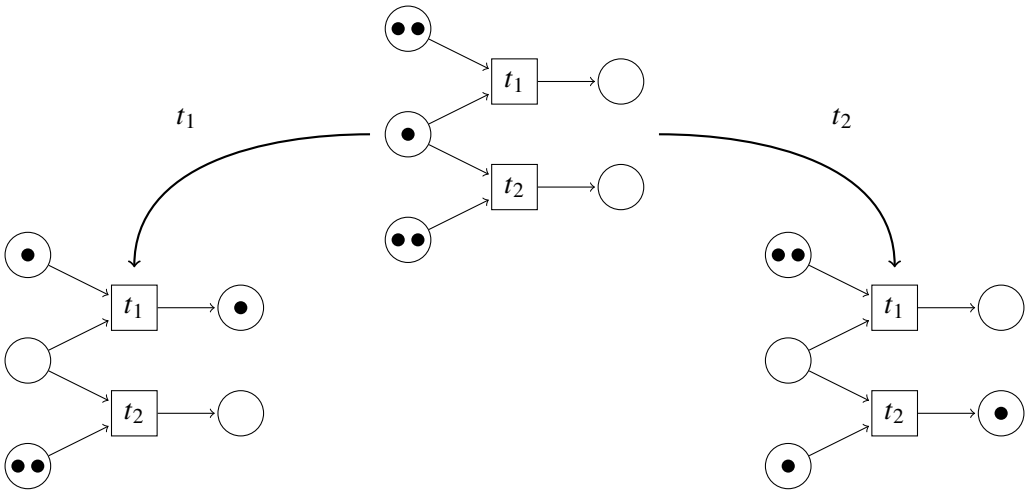
\begin{figure}[htb]
  \centering
  \begin{tikzpicture}[node distance=1.4cm]
    \begin{scope}
      \node [place,tokens=2] (i01) {};
      \node [place,tokens=1] (i02) [below of=i01] {};
      \node [place,tokens=2] (i03) [below of=i02] {};

      \node [transition] (t01) [right of=i01,yshift=-0.7cm] {$t_1$}
        edge [pre] (i01)
        edge [pre] (i02);
      \node [transition] (t02) [below of=t01] {$t_2$}
        edge [pre] (i02)
        edge [pre] (i03);

      \node [place] (o01) [right of=t01] {} edge [pre] (t01);
      \node [place] (o02) [right of=t02] {} edge [pre] (t02);
    \end{scope}

    \begin{scope}[xshift=-5cm,yshift=-3cm]
      \node [place,tokens=1] (i11) {};
      \node [place] (i12) [below of=i11] {};
      \node [place,tokens=2] (i13) [below of=i12] {};

      \node [transition] (t11) [right of=i11,yshift=-0.7cm] {$t_1$}
        edge [pre] (i11)
        edge [pre] (i12);
      \node [transition] (t12) [below of=t11] {$t_2$}
        edge [pre] (i12)
        edge [pre] (i13);

      \node [place,tokens=1] (o11) [right of=t11] {} edge [pre] (t11);
      \node [place] (o12) [right of=t12] {} edge [pre] (t12);
    \end{scope}

    \begin{scope}[xshift=5cm,yshift=-3cm]
      \node [place,tokens=2] (i21) {};
      \node [place] (i22) [below of=i21] {};
      \node [place,tokens=1] (i23) [below of=i22] {};

      \node [transition] (t21) [right of=i21,yshift=-0.7cm] {$t_1$}
        edge [pre] (i21)
        edge [pre] (i22);
      \node [transition] (t22) [below of=t21] {$t_2$}
        edge [pre] (i22)
        edge [pre] (i23);

      \node [place] (o21) [right of=t21] {} edge [pre] (t21);
      \node [place,tokens=1] (o22) [right of=t22] {} edge [pre] (t22);
    \end{scope}

    \path ([xshift=-5mm]i02|-i02) edge [thick,post,out=180,in=90]
      node [pos=0.6,above=0.5cm] {$t_1$} ([yshift=5mm]t11|-t11);
    \path ([xshift=5mm]i02-|o01) edge [thick,post,out=0,in=90]
      node [pos=0.6,above=0.5cm] {$t_2$} ([yshift=5mm]t21|-t21);
  \end{tikzpicture}
  \caption{Conflict of two transitions leading to different program flows.
  }
  \label{fig:conflict}
\end{figure}

In the basic formulation of a Petri net, tokens are structureless and not distinguishable. In order to make programming in terms of Petri nets a practically feasible task, Petri nets in \GPISpace are augmented with additional features:
\begin{enumerate}
  \item The most important addition is for tokens to hold actual data (which corresponds to the theoretical concept of so-called \textit{colored} Petri nets). \GPISpace offers primitive data types and allows the use of user-defined types via an embedded programming language. In this way it possible to impose additional restrictions on the tokens which a transition can accept. This can be used to implement \texttt{if}-\texttt{then}-\texttt{else} and loop functionality on the Petri net level. In particular, it is then possible to connect a place to more than one transition as an input without ambiguity (a token might otherwise be consumed by any connected transition at random, see for example Figure~\ref{fig:conflict}). 
  In fact, even though \GPISpace allows that \textit{conflicts} may arise and then randomly selects a transition to fire, in almost all cases Petri nets are easier to maintain if there never exist more than one transition which can consume a given token.
  \item In colored Petri nets, tokens can be equipped with a type and transitions may impose type restrictions on their input. In \GPISpace, places and transitions are strictly typed, which means that they must be defined with a fixed type signature. In particular, places and transitions can only be connected if these signatures match.
  \item For small computations, an embedded programming language can be used to carry out so-called \textit{tiny computations}, which are not scheduled for execution in a (potentially remote) worker process, but are instead executed directly in the workflow engine. For tasks involving only small arithmetic computations or manipulation of container data, this avoids unnecessary overhead.
\end{enumerate}

The execution of a Petri net is non-deterministic in both the choice of which transition to fire and the choice which token to consume. The actual choice depends on a random number generator and variations in the execution times of the transitions. In other words, in order to facilitate parallelism, \textsc{GPI-Space} forces the applications to use a functional approach to programming which does not depend on an external state. Especially in the context of computer algebra, the introduction of non-determinism has proved to lead to an increase in consistency and predictability of our algorithms.

\subsection{The Petri Net for the Matrix Reduction of a Single Cut}

The matrix reduction performed on a given set of IBP identities is the most involved part in terms of computation time and memory, which makes it the most suitable component of our algorithm for a massively parallel implementation. As previously described, this is achieved by substituting a subset of parameters with a number of numeric values, carrying out the necessary computations in the semi-numeric setting, and then recombining the results into the final, row-reduced linear system via polynomial interpolation.

\GPISpace allows the user to attach so-called \textit{plugins} to a workflow,  which can insert tokens (holding data) into the net, independently of the execution of the Petri net by firing transitions. Compared to traditional Petri net formulations, one main advantage of this plugin system is the ability to maintain a \textit{global state} of the computation: Emulating such a state as a single token in a Petri net would require a complete rewrite for each update to the state, since a transition would first have to consume this token and then place back the modified token. For larger tokens, this introduces an unreasonably large overhead and may impact the performance significantly. In the initial version of our IBP reduction Petri net, this was circumvented by simply starting a remote process which kept track of the state. Where required, a transition could then communicate with this process to send and receive tokens.


Using the plugin system, this approach can now be realized without running processes outside of the execution of \GPISpace. The plugin-managed state is created and controlled by the Petri net and its GPI-Space provided identifier is used by the Petri net to steer all modifications and retrieve information stored in the plugin-managed state. The Petri net uses (specifically annotated) transitions to transport tokens into the plugin. The plugin-transition can receive tokens from the plugin and insert them into the Petri net just like every "normal" transition. In addition, the plugin can also work independently from and in parallel to the Petri net and use (specifically annotated) places to \textit{inject} tokens into the running Petri net. In summary: The Petri net plugins are a mechanism to enable Petri nets to explicitly manage \textit{mutable} state without involving the scheduler and the runtime system, while retaining the conceptual advantages of the Petri net formalism.

The Petri net shown in Figure~\ref{fig:fullnet:new} illustrates the current version of the Petri for computing the matrix reduction of the IBP identities obtained from a single cut.  Dashed arrows refer to \textit{read-only} connections, which means that tokens are in fact not consumed, allowing parallel access to data where modification is not necessary. Dotted arrows show the special data and token movement between the net and the plugin. Finally, transitions may be annotated with conditions restricting their execution, which enables us to implement conditional execution on the Petri net level. In the following, we describe the execution structure of the net:

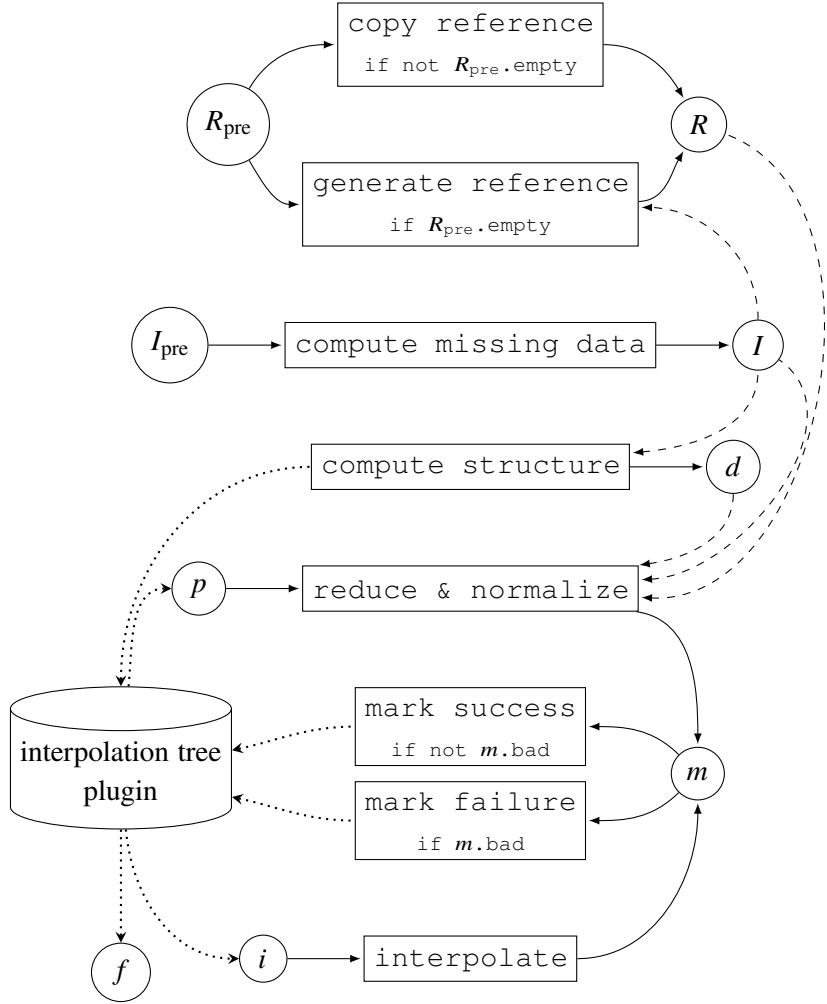
\begin{figure}[tbhp]
  \centering
  \begin{tikzpicture}
    \tikzset{>=latex}
    \tikzset{transition/.append style={font=\ttfamily}}
    \node[transition, align=center] at (0,0)
      (cpy-ref) {copy reference\\\scriptsize{if not $R_{\text{pre}}$.empty}};
    \node[transition, align=center, below=of cpy-ref]
      (gen-ref) {generate reference\\\scriptsize{if $R_{\text{pre}}$.empty}};
    \path (gen-ref) -- (cpy-ref) node[midway] (ref-mid) {};

    \node[place,left=2.5 of ref-mid] (preref) {$R_{\text{pre}}$}
      edge[post, out=-60, in=180] (gen-ref)
      edge[post, out=60, in=180] (cpy-ref);
    \node[place,right=2.5 of ref-mid] (ref) {$R$}
      edge[pre, out=-120, in=1] (gen-ref)
      edge[pre, out=120, in=0] (cpy-ref);

    \node[transition, below=of gen-ref] (comp-miss) {compute missing data};
    \node[place, left=of comp-miss] (preinput) {$I_{\text{pre}}$}
      edge[post] (comp-miss);
    \node[place, right=of comp-miss] (input) {$I$}
      edge[pre] (comp-miss)
      edge[post,dashed,out=90,in=-1] (gen-ref);

    \node[transition, below=of comp-miss] (comp-struct) {compute structure}
      edge[pre,dashed,out=5,in=-90] (input);
    \node[place,right=of comp-struct] (data) {$d$}
      edge[pre] (comp-struct);

    \node[transition,below=of comp-struct] (reduce) {reduce \& normalize}
      edge[pre,dashed,out=-3,in=-20] (ref)
      edge[pre,dashed,out=3,in=-35] (input)
      edge[pre,dashed,out=8,in=south] (data);
    \node[place, left=of reduce] (pts) {$p$}
      edge[post] (reduce);
    \node[transition,below=of reduce,align=center]
      (mark-succ) {mark success\\\scriptsize{if not $m$.bad}};
    \node[transition,below=0.2 of mark-succ,align=center]
      (mark-fail) {mark failure\\\scriptsize{if $m$.bad}};
    \path (mark-succ) -- (mark-fail) node[midway] (mark-mid) {};
    \node[transition,below=of mark-fail] (interpolate) {interpolate};
    \node[place,left=of interpolate] (ip) {$i$}
      edge[post] (interpolate);
    \node[place, right=2.5 of mark-mid] (rref) {$m$}
      edge[post,out=135,in=east] (mark-succ)
      edge[post,out=-135,in=east] (mark-fail)
      edge[pre,out=north,in=-8] (reduce)
      edge[pre,out=south,in=east] (interpolate);

    \node[database,left=3 of mark-mid,align=center]
      (stor) {interpolation tree\\ plugin}
      edge[stealth-,thick,dotted,out=12,in=west] (mark-succ)
      edge[stealth-,thick,dotted,out=-12,in=west] (mark-fail)
      edge[-stealth,thick,dotted,out=85,in=west] (pts)
      edge[stealth-,thick,dotted,out=north,in=west] (comp-struct)
      edge[-stealth,thick,dotted,out=-85,in=west] (ip);

    \node[place,below=1.5 of stor] (fin) {$f$}
      edge[stealth-,thick,dotted] (stor);
  \end{tikzpicture}
  \caption{Updated and simplified Petri net incorporating plugin system.}
  \label{fig:fullnet:new}
\end{figure}

\begin{description}
  \item[Input tokens:] The net is initialized with two tokens. One is the input data token which is put onto $I_{\text{pre}}$. It is a structured type with the following fields:
    \begin{itemize}
      \item The (filename pointing to the) input linear relations stored as a matrix over a rational function field $\setQ(\vec c, D)$.
      \item The list of indices of the parameters which will be substituted during the reduction computation and interpolated, say $\{1, \dots, r\}$.
      \item The list of indices of parameters which will be substituted during the reference reduction. This is not necessarily the complement of the set of interpolated parameters, for instance, if some parameters in the function field are of no interest during the reduction computations.
      \item The list of target integral indices.
      \item Optionally: A precomputed trace for row and column swaps during the reduction step. This field will be referred to as \texttt{trace} in the Petri net.
      \item Optionally: A list of master integral indices. These indices can be obtained by a fixed-column computation and thus can be omitted. In the net, this field is referred to as \texttt{master}.
    \end{itemize}
    The second token is a reference token which is required by our rational function interpolation algorithm and contains an  result for the complementary set of analytic parameters, that is, exchanging in the hybrid approach the interpolated and the analytic parameters. This token is put onto $R_{\text{pre}}$. This token optionally contains a reference substitution point and the relations from $I_\text{pre}$ already reduced according to this point. The data field \texttt{empty} of the token can be used to indicate that this token does \textit{not} contain this information.

    This option is useful in the following sense: If the number of substituted parameters is considerably larger than the set of reference parameters, then a direct computation of the reference might not be feasible. In this case, the reference is computed by iteratively using the parallel algorithm.
  \item[Transition \texttt{copy reference}:] If for a token on the place $R_{\text{pre}}$ the field \texttt{$R_{\text{pre}}$.empty} has the value false, this token holds a precomputed reference matrix. In this case, the token is simply moved to $R$ by the transition \texttt{copy reference}.
  \item[Transition \texttt{generate reference}:] Otherwise, that is, if \texttt{$R_{\text{pre}}$.empty} is true, no reference matrix was supplied. This transition then generates a random reference point and computes the reference matrix, which is then put onto the place $R$.
  \item[Transition \texttt{compute missing data}:] This transition computes the optional fields \texttt{trace} and \texttt{master}, if they are missing from the input token on $I_{\text{pre}}$. The completed input token is then put onto the place $I$. In practice, this transition is a slightly more involved subnet. Since it is functionally and structurally not particularly interesting, it is omitted from our presentation.
  \item[Transition \texttt{compute structure}:] Again, this is a simplification of a more involved subnet to simplify the presentation. Here, the polynomial degrees of all coefficients of the fully reduced linear system are computed for each parameter via univariate computations. The maximum degrees per parameter are supplied to the interpolation tree plugin to determine the required number of interpolation points. A list of matrices which hold the degrees of the numerator and denominator of each matrix entry is then put onto the place $d$. This data on the polynomial degrees is required in the normalization step.
  \item[Interpolation Tree Plugin:] The tree manages the interpolation progress in a tree structure, as outlined above. As soon as the degree data is determined, the plugin can generate an initial required set of interpolation points, which are then put onto the place $p$. If reductions fail or result in bad data, points are marked as failures in the tree. Accordingly, the plugin generates new points to ensure that sufficient data for the interpolation is available. If, however, enough reductions (or interpolations) for a given parameter are successful, the plugin generates the corresponding point for this interpolation, which is then put onto the place $i$. When the root of the interpolation tree, which corresponds to the fully interpolated result, is marked as completed, the interpolation tree plugin outputs the filename of the final result as a token onto the place $f$. This terminates the execution of the Petri net.
  \item[Transition \texttt{reduce \& normalize}:] For a point token from $p$, this transition performs the matrix reduction according to the traces supplied in the input token on $I$. As we perform division during the computation with fixed column and row swap traces, we may encounter division by zero after substituting certain integer values. In this case, the output token will be marked as \textit{bad}. With the degree data from $d$ and the reference reduction result from $R$, the transition in addition normalizes the successful reduction result and potentially marks it as \textit{bad}, if polynomial cancellation occurred in some entry. Output tokens store the associated substitution point and (the filename of) the reduction and normalization result if it exists. Accordingly, the token has the additional field \texttt{bad} and is put onto the place $m$.
  \item[Transitions \texttt{mark success} and \texttt{mark failure}:] As stated above, the tokens on $m$ indicate with the field \texttt{bad} whether the reduction and normalization computations produced a valid result. These two transitions then mark the associated substitution point accordingly as \textit{completed} or \textit{failed} in the interpolation tree.
  \item[Transition \texttt{interpolate}:] The tokens in $i$ are generated by the interpolation plugin to contain the filenames for all matrices required for the particular interpolation. This transition then simply computes the polynomial interpolations for both numerators and denominators of the result matrix. The resulting matrix and the substitution point which correspond to this interpolation are then put onto the place $m$. These tokens will always have a \texttt{bad}-value of false.
\end{description}

As stated in the description above, the Petri net execution terminates as soon as the output matrix is fully interpolated and the corresponding token is put onto the place $f$.

A desirable feature of long-running large-scale computations is restartability from a previously reached state, due to the possibility of compute node failures or job cancellations upon reaching imposed time limits. For our algorithm, this can be achieved by storing -- or recreating -- the interpolation tree managed by the plugin. Since the tree creation deterministically depends on the list of parameter indices and their degrees, recreating subsequent states can be achieved by logging the \textit{modifications} of this tree, that is, by recording whether a node was marked as failed or successful, and re-applying them. Since the information logged in this way does not comprise the full structure for most of the Petri net execution, this is more efficient than storing the whole tree. Of course, this requires the user to restart the algorithm without changing its parameters, which introduces potentially undesirable inflexibility.

\section{Example}
\input{Examples.tex}
\section{Summary and Outlook}
We have illustrated some recent progress that has been made on tackling the IBP reduction problem. In our approach, the problem is simplified in different algorithmic sub-steps: Firstly by using methods from algebraic geometry to simplify the generated IBP system. A Mathematica interface for generating IBPs with restricted propagator degrees is presented. Secondly by using the \textsc{Singular}-\textsc{GPI-Space} framework to perform Gaussian elimination relying on interpolation of parameters in a massively parallel way. Finally, we have seen how a good choice of basis, in our case the dlog basis, and an efficient algebraic representation of the data can significantly simplify the output of the reduction.

With these new developments, we could hope to tackle higher points or higher loop reductions and amplitudes with increasing number of off-shell external points or internal masses, thus increasing the precision of the theoretical predictions for processes of phenomenological interest.
Moreover, we could apply the \textsc{Singular}-\textsc{GPI-Space} framework to other problems than IBP reduction, for example, in recent years some progress has been made into studying the Bethe ansatz equations and partition functions of integrable models \cite{Bajnok:2020xoz} or for the study of Grassmanians and their tropical varieties \cite{2020arXiv200313752B}.

\bibliographystyle{JHEP}
\bibliography{proceeding.bib}

\providecommand{\href}[2]{#2}\begingroup\raggedright\begin{thebibliography}{10}

\bibitem{Badger:2013gxa}
S.~Badger, H.~Frellesvig, and Y.~Zhang, {\it {A Two-Loop Five-Gluon Helicity
  Amplitude in QCD}},  {\em JHEP} {\bf 12} (2013) 045,
  [\href{http://arxiv.org/abs/1310.1051}{{\tt arXiv:1310.1051}}].

\bibitem{Gehrmann:2015bfy}
T.~Gehrmann, J.~M. Henn, and N.~A. Lo~Presti, {\it {Analytic form of the
  two-loop planar five-gluon all-plus-helicity amplitude in QCD}},  {\em Phys.
  Rev. Lett.} {\bf 116} (2016), no.~6 062001,
  [\href{http://arxiv.org/abs/1511.05409}{{\tt arXiv:1511.05409}}]. [Erratum:
  Phys. Rev. Lett.116,no.18,189903(2016)].

\bibitem{Badger:2017jhb}
S.~Badger, C.~Br{\o}nnum-Hansen, H.~B. Hartanto, and T.~Peraro, {\it {First
  look at two-loop five-gluon scattering in QCD}},  {\em Phys. Rev. Lett.} {\bf
  120} (2018), no.~9 092001, [\href{http://arxiv.org/abs/1712.02229}{{\tt
  arXiv:1712.02229}}].

\bibitem{Abreu:2017hqn}
S.~Abreu, F.~Febres~Cordero, H.~Ita, B.~Page, and M.~Zeng, {\it {Planar
  Two-Loop Five-Gluon Amplitudes from Numerical Unitarity}},  {\em Phys. Rev.}
  {\bf D97} (2018), no.~11 116014, [\href{http://arxiv.org/abs/1712.03946}{{\tt
  arXiv:1712.03946}}].

\bibitem{Abreu:2018aqd}
S.~Abreu, L.~J. Dixon, E.~Herrmann, B.~Page, and M.~Zeng, {\it {The two-loop
  five-point amplitude in $\mathcal{N} =4$ super-Yang-Mills theory}},  {\em
  Phys. Rev. Lett.} {\bf 122} (2019), no.~12 121603,
  [\href{http://arxiv.org/abs/1812.08941}{{\tt arXiv:1812.08941}}].

\bibitem{Abreu:2018jgq}
S.~Abreu, F.~Febres~Cordero, H.~Ita, B.~Page, and V.~Sotnikov, {\it {Planar
  Two-Loop Five-Parton Amplitudes from Numerical Unitarity}},  {\em JHEP} {\bf
  11} (2018) 116, [\href{http://arxiv.org/abs/1809.09067}{{\tt
  arXiv:1809.09067}}].

\bibitem{Boels:2018nrr}
R.~H. Boels, Q.~Jin, and H.~Luo, {\it {Efficient integrand reduction for
  particles with spin}},  \href{http://arxiv.org/abs/1802.06761}{{\tt
  arXiv:1802.06761}}.

\bibitem{Gehrmann:2018yef}
T.~Gehrmann, J.~M. Henn, and N.~A. Lo~Presti, {\it {Pentagon functions for
  massless planar scattering amplitudes}},  {\em JHEP} {\bf 10} (2018) 103,
  [\href{http://arxiv.org/abs/1807.09812}{{\tt arXiv:1807.09812}}].

\bibitem{Badger:2018enw}
S.~Badger, C.~Br{\o}nnum-Hansen, H.~B. Hartanto, and T.~Peraro, {\it {Analytic
  helicity amplitudes for two-loop five-gluon scattering: the single-minus
  case}},  {\em JHEP} {\bf 01} (2019) 186,
  [\href{http://arxiv.org/abs/1811.11699}{{\tt arXiv:1811.11699}}].

\bibitem{Abreu:2018zmy}
S.~Abreu, J.~Dormans, F.~Febres~Cordero, H.~Ita, and B.~Page, {\it {Analytic
  Form of Planar Two-Loop Five-Gluon Scattering Amplitudes in QCD}},  {\em
  Phys. Rev. Lett.} {\bf 122} (2019), no.~8 082002,
  [\href{http://arxiv.org/abs/1812.04586}{{\tt arXiv:1812.04586}}].

\bibitem{Chicherin:2018yne}
D.~Chicherin, T.~Gehrmann, J.~M. Henn, P.~Wasser, Y.~Zhang, and S.~Zoia, {\it
  {Analytic result for a two-loop five-particle amplitude}},  {\em Phys. Rev.
  Lett.} {\bf 122} (2019), no.~12 121602,
  [\href{http://arxiv.org/abs/1812.11057}{{\tt arXiv:1812.11057}}].

\bibitem{Chicherin:2019xeg}
D.~Chicherin, T.~Gehrmann, J.~M. Henn, P.~Wasser, Y.~Zhang, and S.~Zoia, {\it
  {The two-loop five-particle amplitude in $ \mathcal{N} $ = 8 supergravity}},
  {\em JHEP} {\bf 03} (2019) 115, [\href{http://arxiv.org/abs/1901.05932}{{\tt
  arXiv:1901.05932}}].

\bibitem{Abreu:2019rpt}
S.~Abreu, L.~J. Dixon, E.~Herrmann, B.~Page, and M.~Zeng, {\it {The two-loop
  five-point amplitude in $ \mathcal{N} $ = 8 supergravity}},  {\em JHEP} {\bf
  03} (2019) 123, [\href{http://arxiv.org/abs/1901.08563}{{\tt
  arXiv:1901.08563}}].

\bibitem{Abreu:2019odu}
S.~Abreu, J.~Dormans, F.~Febres~Cordero, H.~Ita, B.~Page, and V.~Sotnikov, {\it
  {Analytic Form of the Planar Two-Loop Five-Parton Scattering Amplitudes in
  QCD}},  {\em JHEP} {\bf 05} (2019) 084,
  [\href{http://arxiv.org/abs/1904.00945}{{\tt arXiv:1904.00945}}].

\bibitem{Hartanto:2019uvl}
H.~B. Hartanto, S.~Badger, C.~Br{\o}nnum-Hansen, and T.~Peraro, {\it {A
  numerical evaluation of planar two-loop helicity amplitudes for a W-boson
  plus four partons}},  \href{http://arxiv.org/abs/1906.11862}{{\tt
  arXiv:1906.11862}}.

\bibitem{CHETYRKIN1981159}
K.~Chetyrkin and F.~Tkachov, {\it Integration by parts: The algorithm to
  calculate $\beta$-functions in 4 loops},  {\em Nuclear Physics B} {\bf 192}
  (1981), no.~1 159 -- 204.

\bibitem{Kotikov:1990kg}
A.~V. Kotikov, {\it {Differential equations method: New technique for massive
  Feynman diagrams calculation}},  {\em Phys. Lett.} {\bf B254} (1991)
  158--164.

\bibitem{Kotikov:1991pm}
A.~V. Kotikov, {\it {Differential equation method: The Calculation of N point
  Feynman diagrams}},  {\em Phys. Lett.} {\bf B267} (1991) 123--127.

\bibitem{Bern:1993kr}
Z.~Bern, L.~J. Dixon, and D.~A. Kosower, {\it {Dimensionally regulated pentagon
  integrals}},  {\em Nucl. Phys.} {\bf B412} (1994) 751--816,
  [\href{http://arxiv.org/abs/hep-ph/9306240}{{\tt hep-ph/9306240}}].

\bibitem{Remiddi:1997ny}
E.~Remiddi, {\it {Differential equations for Feynman graph amplitudes}},  {\em
  Nuovo Cim.} {\bf A110} (1997) 1435--1452,
  [\href{http://arxiv.org/abs/hep-th/9711188}{{\tt hep-th/9711188}}].

\bibitem{Gehrmann:1999as}
T.~Gehrmann and E.~Remiddi, {\it {Differential equations for two loop four
  point functions}},  {\em Nucl. Phys.} {\bf B580} (2000) 485--518,
  [\href{http://arxiv.org/abs/hep-ph/9912329}{{\tt hep-ph/9912329}}].

\bibitem{Henn:2013pwa}
J.~M. Henn, {\it {Multiloop integrals in dimensional regularization made
  simple}},  {\em Phys. Rev. Lett.} {\bf 110} (2013) 251601,
  [\href{http://arxiv.org/abs/1304.1806}{{\tt arXiv:1304.1806}}].

\bibitem{Papadopoulos:2014lla}
C.~G. Papadopoulos, {\it {Simplified differential equations approach for Master
  Integrals}},  {\em JHEP} {\bf 07} (2014) 088,
  [\href{http://arxiv.org/abs/1401.6057}{{\tt arXiv:1401.6057}}].

\bibitem{Lee:2014ioa}
R.~N. Lee, {\it {Reducing differential equations for multiloop master
  integrals}},  {\em JHEP} {\bf 04} (2015) 108,
  [\href{http://arxiv.org/abs/1411.0911}{{\tt arXiv:1411.0911}}].

\bibitem{Ablinger:2015tua}
J.~Ablinger, A.~Behring, J.~Bl{\"u}mlein, A.~De~Freitas, A.~von Manteuffel, and
  C.~Schneider, {\it {Calculating Three Loop Ladder and V-Topologies for
  Massive Operator Matrix Elements by Computer Algebra}},  {\em Comput. Phys.
  Commun.} {\bf 202} (2016) 33--112,
  [\href{http://arxiv.org/abs/1509.08324}{{\tt arXiv:1509.08324}}].

\bibitem{Papadopoulos:2015jft}
C.~G. Papadopoulos, D.~Tommasini, and C.~Wever, {\it {The Pentabox Master
  Integrals with the Simplified Differential Equations approach}},  {\em JHEP}
  {\bf 04} (2016) 078, [\href{http://arxiv.org/abs/1511.09404}{{\tt
  arXiv:1511.09404}}].

\bibitem{Liu:2017jxz}
X.~Liu, Y.-Q. Ma, and C.-Y. Wang, {\it {A Systematic and Efficient Method to
  Compute Multi-loop Master Integrals}},  {\em Phys. Lett.} {\bf B779} (2018)
  353--357, [\href{http://arxiv.org/abs/1711.09572}{{\tt arXiv:1711.09572}}].

\bibitem{Dixon:2011pw}
L.~J. Dixon, J.~M. Drummond, and J.~M. Henn, {\it {Bootstrapping the three-loop
  hexagon}},  {\em JHEP} {\bf 11} (2011) 023,
  [\href{http://arxiv.org/abs/1108.4461}{{\tt arXiv:1108.4461}}].

\bibitem{Dixon:2013eka}
L.~J. Dixon, J.~M. Drummond, M.~von Hippel, and J.~Pennington, {\it {Hexagon
  functions and the three-loop remainder function}},  {\em JHEP} {\bf 12}
  (2013) 049, [\href{http://arxiv.org/abs/1308.2276}{{\tt arXiv:1308.2276}}].

\bibitem{Dixon:2014iba}
L.~J. Dixon and M.~von Hippel, {\it {Bootstrapping an NMHV amplitude through
  three loops}},  {\em JHEP} {\bf 10} (2014) 065,
  [\href{http://arxiv.org/abs/1408.1505}{{\tt arXiv:1408.1505}}].

\bibitem{Caron-Huot:2016owq}
S.~Caron-Huot, L.~J. Dixon, A.~McLeod, and M.~von Hippel, {\it {Bootstrapping a
  Five-Loop Amplitude Using Steinmann Relations}},  {\em Phys. Rev. Lett.} {\bf
  117} (2016), no.~24 241601, [\href{http://arxiv.org/abs/1609.00669}{{\tt
  arXiv:1609.00669}}].

\bibitem{Dixon:2015iva}
L.~J. Dixon, M.~von Hippel, and A.~J. McLeod, {\it {The four-loop six-gluon
  NMHV ratio function}},  {\em JHEP} {\bf 01} (2016) 053,
  [\href{http://arxiv.org/abs/1509.08127}{{\tt arXiv:1509.08127}}].

\bibitem{Dixon:2016nkn}
L.~J. Dixon, J.~Drummond, T.~Harrington, A.~J. McLeod, G.~Papathanasiou, and
  M.~Spradlin, {\it {Heptagons from the Steinmann Cluster Bootstrap}},  {\em
  JHEP} {\bf 02} (2017) 137, [\href{http://arxiv.org/abs/1612.08976}{{\tt
  arXiv:1612.08976}}].

\bibitem{Chicherin:2017dob}
D.~Chicherin, J.~Henn, and V.~Mitev, {\it {Bootstrapping pentagon functions}},
  {\em JHEP} {\bf 05} (2018) 164, [\href{http://arxiv.org/abs/1712.09610}{{\tt
  arXiv:1712.09610}}].

\bibitem{Caron-Huot:2019vjl}
S.~Caron-Huot, L.~J. Dixon, F.~Dulat, M.~von Hippel, A.~J. McLeod, and
  G.~Papathanasiou, {\it {Six-Gluon Amplitudes in Planar ${\cal N}=4$
  Super-Yang-Mills Theory at Six and Seven Loops}},
  \href{http://arxiv.org/abs/1903.10890}{{\tt arXiv:1903.10890}}.

\bibitem{Ita:2015tya}
H.~Ita, {\it {Two-loop Integrand Decomposition into Master Integrals and
  Surface Terms}},  {\em Phys. Rev.} {\bf D94} (2016), no.~11 116015,
  [\href{http://arxiv.org/abs/1510.05626}{{\tt arXiv:1510.05626}}].

\bibitem{Abreu:2017xsl}
S.~Abreu, F.~Febres~Cordero, H.~Ita, M.~Jaquier, B.~Page, and M.~Zeng, {\it
  {Two-Loop Four-Gluon Amplitudes from Numerical Unitarity}},  {\em Phys. Rev.
  Lett.} {\bf 119} (2017), no.~14 142001,
  [\href{http://arxiv.org/abs/1703.05273}{{\tt arXiv:1703.05273}}].

\bibitem{Badger:2019djh}
S.~Badger, D.~Chicherin, T.~Gehrmann, G.~Heinrich, J.~M. Henn, T.~Peraro,
  P.~Wasser, Y.~Zhang, and S.~Zoia, {\it {Analytic form of the full two-loop
  five-gluon all-plus helicity amplitude}},  {\em Phys. Rev. Lett.} {\bf 123}
  (2019), no.~7 071601, [\href{http://arxiv.org/abs/1905.03733}{{\tt
  arXiv:1905.03733}}].

\bibitem{Mastrolia:2018uzb}
P.~Mastrolia and S.~Mizera, {\it {Feynman Integrals and Intersection Theory}},
  {\em JHEP} {\bf 02} (2019) 139, [\href{http://arxiv.org/abs/1810.03818}{{\tt
  arXiv:1810.03818}}].

\bibitem{Frellesvig:2019uqt}
H.~Frellesvig, F.~Gasparotto, M.~K. Mandal, P.~Mastrolia, L.~Mattiazzi, and
  S.~Mizera, {\it {Vector Space of Feynman Integrals and Multivariate
  Intersection Numbers}},  \href{http://arxiv.org/abs/1907.02000}{{\tt
  arXiv:1907.02000}}.

\bibitem{Frellesvig:2019kgj}
H.~Frellesvig, F.~Gasparotto, S.~Laporta, M.~K. Mandal, P.~Mastrolia,
  L.~Mattiazzi, and S.~Mizera, {\it {Decomposition of Feynman Integrals on the
  Maximal Cut by Intersection Numbers}},  {\em JHEP} {\bf 05} (2019) 153,
  [\href{http://arxiv.org/abs/1901.11510}{{\tt arXiv:1901.11510}}].

\bibitem{Frellesvig:2020qot}
H.~Frellesvig, F.~Gasparotto, S.~Laporta, M.~K. Mandal, P.~Mastrolia,
  L.~Mattiazzi, and S.~Mizera, {\it {Decomposition of Feynman Integrals by
  Multivariate Intersection Numbers}},
  \href{http://arxiv.org/abs/2008.04823}{{\tt arXiv:2008.04823}}.

\bibitem{Liu:2018dmc}
X.~Liu and Y.-Q. Ma, {\it {Determining arbitrary Feynman integrals by vacuum
  integrals}},  {\em Phys. Rev.} {\bf D99} (2019), no.~7 071501,
  [\href{http://arxiv.org/abs/1801.10523}{{\tt arXiv:1801.10523}}].

\bibitem{Guan:2019bcx}
X.~Guan, X.~Liu, and Y.-Q. Ma, {\it {Complete reduction of two-loop
  five-light-parton scattering amplitudes}},
  \href{http://arxiv.org/abs/1912.09294}{{\tt arXiv:1912.09294}}.

\bibitem{Zhang:2018mlo}
P.~Zhang, C.-Y. Wang, X.~Liu, Y.-Q. Ma, C.~Meng, and K.-T. Chao, {\it
  {Semi-analytical calculation of gluon fragmentation into$^{1}$S$_{0}^{[1,8]}$
  quarkonia at next-to-leading order}},  {\em JHEP} {\bf 04} (2019) 116,
  [\href{http://arxiv.org/abs/1810.07656}{{\tt arXiv:1810.07656}}].

\bibitem{Wang:2019mnn}
Y.~Wang, Z.~Li, and N.~Ul~Basat, {\it {Direct Reduction of Amplitude}},
  \href{http://arxiv.org/abs/1901.09390}{{\tt arXiv:1901.09390}}.

\bibitem{Laporta:2001dd}
S.~Laporta, {\it {High precision calculation of multiloop Feynman integrals by
  difference equations}},  {\em Int. J. Mod. Phys.} {\bf A15} (2000)
  5087--5159, [\href{http://arxiv.org/abs/hep-ph/0102033}{{\tt
  hep-ph/0102033}}].

\bibitem{Smirnov:2008iw}
A.~V. Smirnov, {\it {Algorithm FIRE -- Feynman Integral REduction}},  {\em
  JHEP} {\bf 10} (2008) 107, [\href{http://arxiv.org/abs/0807.3243}{{\tt
  arXiv:0807.3243}}].

\bibitem{Smirnov:2014hma}
A.~V. Smirnov, {\it {FIRE5: a C++ implementation of Feynman Integral
  REduction}},  {\em Comput. Phys. Commun.} {\bf 189} (2015) 182--191,
  [\href{http://arxiv.org/abs/1408.2372}{{\tt arXiv:1408.2372}}].

\bibitem{Smirnov:2019qkx}
A.~V. Smirnov and F.~S. Chuharev, {\it {FIRE6: Feynman Integral REduction with
  Modular Arithmetic}},  \href{http://arxiv.org/abs/1901.07808}{{\tt
  arXiv:1901.07808}}.

\bibitem{Maierhoefer:2017hyi}
P.~Maierhoefer, J.~Usovitsch, and P.~Uwer, {\it {Kira - A Feynman Integral
  Reduction Program}},  {\em Comput. Phys. Commun.} {\bf 230} (2018) 99--112,
  [\href{http://arxiv.org/abs/1705.05610}{{\tt arXiv:1705.05610}}].

\bibitem{Maierhofer:2018gpa}
P.~Maierh{\"o}fer and J.~Usovitsch, {\it {Kira 1.2 Release Notes}},
  \href{http://arxiv.org/abs/1812.01491}{{\tt arXiv:1812.01491}}.

\bibitem{vonManteuffel:2012np}
A.~von Manteuffel and C.~Studerus, {\it {Reduze 2 - Distributed Feynman
  Integral Reduction}},  \href{http://arxiv.org/abs/1201.4330}{{\tt
  arXiv:1201.4330}}.

\bibitem{Klappert:2020nbg}
J.~Klappert, F.~Lange, P.~Maierh\"ofer, and J.~Usovitsch, {\it {Integral
  Reduction with Kira 2.0 and Finite Field Methods}},
  \href{http://arxiv.org/abs/2008.06494}{{\tt arXiv:2008.06494}}.

\bibitem{Lee:2013hzt}
R.~N. Lee and A.~A. Pomeransky, {\it {Critical points and number of master
  integrals}},  {\em JHEP} {\bf 11} (2013) 165,
  [\href{http://arxiv.org/abs/1308.6676}{{\tt arXiv:1308.6676}}].

\bibitem{Georgoudis:2016wff}
A.~Georgoudis, K.~J. Larsen, and Y.~Zhang, {\it {Azurite: An algebraic geometry
  based package for finding bases of loop integrals}},  {\em Comput. Phys.
  Commun.} {\bf 221} (2017) 203--215,
  [\href{http://arxiv.org/abs/1612.04252}{{\tt arXiv:1612.04252}}].

\bibitem{Bendle:2019csk}
D.~Bendle, J.~B\"ohm, W.~Decker, A.~Georgoudis, F.-J. Pfreundt, M.~Rahn,
  P.~Wasser, and Y.~Zhang, {\it {Integration-by-parts reductions of Feynman
  integrals using Singular and GPI-Space}},  {\em JHEP} {\bf 02} (2020) 079,
  [\href{http://arxiv.org/abs/1908.04301}{{\tt arXiv:1908.04301}}].

\bibitem{Usovitsch:2020jrk}
J.~Usovitsch, {\it {Factorization of denominators in integration-by-parts
  reductions}},  \href{http://arxiv.org/abs/2002.08173}{{\tt
  arXiv:2002.08173}}.

\bibitem{Smirnov:2020quc}
A.~Smirnov and V.~Smirnov, {\it {How to choose master integrals}},
  \href{http://arxiv.org/abs/2002.08042}{{\tt arXiv:2002.08042}}.

\bibitem{Boehm:2020ijp}
J.~B{\"o}hm, M.~Wittmann, Z.~Wu, Y.~Xu, and Y.~Zhang, {\it {IBP reduction
  coefficients made simple}},  \href{http://arxiv.org/abs/2008.13194}{{\tt
  arXiv:2008.13194}}.

\bibitem{Boehm:2017wjc}
J.~B{\"o}hm, A.~Georgoudis, K.~J. Larsen, M.~Schulze, and Y.~Zhang, {\it
  {Complete sets of logarithmic vector fields for integration-by-parts
  identities of Feynman integrals}},  {\em Phys. Rev.} {\bf D98} (2018), no.~2
  025023, [\href{http://arxiv.org/abs/1712.09737}{{\tt arXiv:1712.09737}}].

\bibitem{Gluza:2010ws}
J.~Gluza, K.~Kajda, and D.~A. Kosower, {\it {Towards a Basis for Planar
  Two-Loop Integrals}},  {\em Phys.Rev.} {\bf D83} (2011) 045012,
  [\href{http://arxiv.org/abs/1009.0472}{{\tt arXiv:1009.0472}}].

\bibitem{Schabinger:2011dz}
R.~M. Schabinger, {\it {A New Algorithm For The Generation Of
  Unitarity-Compatible Integration By Parts Relations}},  {\em JHEP} {\bf 01}
  (2012) 077, [\href{http://arxiv.org/abs/1111.4220}{{\tt arXiv:1111.4220}}].

\bibitem{Larsen:2015ped}
K.~J. Larsen and Y.~Zhang, {\it {Integration-by-parts reductions from unitarity
  cuts and algebraic geometry}},  {\em Phys. Rev.} {\bf D93} (2016), no.~4
  041701, [\href{http://arxiv.org/abs/1511.01071}{{\tt arXiv:1511.01071}}].

\bibitem{Boehm:2018fpv}
J.~B{\"o}hm, A.~Georgoudis, K.~J. Larsen, H.~Sch{\"o}nemann, and Y.~Zhang, {\it
  {Complete integration-by-parts reductions of the non-planar hexagon-box via
  module intersections}},  {\em JHEP} {\bf 09} (2018) 024,
  [\href{http://arxiv.org/abs/1805.01873}{{\tt arXiv:1805.01873}}].

\bibitem{Chawdhry:2018awn}
H.~A. Chawdhry, M.~A. Lim, and A.~Mitov, {\it {Two-loop five-point massless QCD
  amplitudes within the IBP approach}},
  \href{http://arxiv.org/abs/1805.09182}{{\tt arXiv:1805.09182}}.

\bibitem{vonManteuffel:2014ixa}
A.~von Manteuffel and R.~M. Schabinger, {\it {A novel approach to integration
  by parts reduction}},  {\em Phys. Lett.} {\bf B744} (2015) 101--104,
  [\href{http://arxiv.org/abs/1406.4513}{{\tt arXiv:1406.4513}}].

\bibitem{Peraro:2016wsq}
T.~Peraro, {\it {Scattering amplitudes over finite fields and multivariate
  functional reconstruction}},  {\em JHEP} {\bf 12} (2016) 030,
  [\href{http://arxiv.org/abs/1608.01902}{{\tt arXiv:1608.01902}}].

\bibitem{Peraro:2019svx}
T.~Peraro, {\it {FiniteFlow: multivariate functional reconstruction using
  finite fields and dataflow graphs}},  {\em JHEP} {\bf 07} (2019) 031,
  [\href{http://arxiv.org/abs/1905.08019}{{\tt arXiv:1905.08019}}].

\bibitem{BDFPRR}
J.~B\"ohm, W.~Decker, A.~Fr\"uhbis-Kr\"uger, F.-J. Pfreundt, M.~Rahn, and
  L.~Ristau, {\it Towards massively parallel computations in algebraic
  geometry},  {\em Foundations of Computational Mathematics} (2020)
  [\href{http://arxiv.org/abs/1808.09727}{{\tt arXiv:1808.09727}}].

\bibitem{singular}
W.~Decker, G.-M. Greuel, G.~Pfister, and H.~Sch\"onemann, ``{\sc Singular}
  {4-1-3} --- {A} computer algebra system for polynomial computations.''
  \href{http://www.singular.uni-kl.de}{http://www.singular.uni-kl.de}, 2020.

\bibitem{GPI}
F.-J. Pfreundt and M.~Rahn, {\it {GPI}-{S}pace},  2018.
\newblock Fraunhofer ITWM Kaiserslautern,
  \href{http://www.gpi-space.de/}{http://www.gpi-space.de/}.

\bibitem{bendle:master}
D.~Bendle, {\it Massively parallel computation of integration-by-parts
  relations for {F}eynman integrals},  2020.
\newblock Master's Thesis.

\bibitem{Chicherin:2018old}
D.~Chicherin, T.~Gehrmann, J.~M. Henn, P.~Wasser, Y.~Zhang, and S.~Zoia, {\it
  {All master integrals for three-jet production at NNLO}},  {\em Phys. Rev.
  Lett.} {\bf 123} (2019), no.~4 041603,
  [\href{http://arxiv.org/abs/1812.11160}{{\tt arXiv:1812.11160}}].

\bibitem{Bajnok:2020xoz}
Z.~Bajnok, J.~L. Jacobsen, Y.~Jiang, R.~I. Nepomechie, and Y.~Zhang, {\it
  {Cylinder partition function of the 6-vertex model from algebraic geometry}},
   {\em JHEP} {\bf 06} (2020) 169, [\href{http://arxiv.org/abs/2002.09019}{{\tt
  arXiv:2002.09019}}].

\bibitem{2020arXiv200313752B}
D.~{Bendle}, J.~B{\"o}hm, Y.~{Ren}, and B.~{Schr{\"o}ter}, {\it {Parallel
  Computation of tropical varieties, their positive part, and tropical
  Grassmannians}},  {\em arXiv e-prints} (Mar., 2020) arXiv:2003.13752,
  [\href{http://arxiv.org/abs/2003.13752}{{\tt arXiv:2003.13752}}].

\end{thebibliography}\endgroup
\end{document}